\pdfoutput = 1

\documentclass[doublecol]{epl2}

\usepackage{color}
\definecolor{greencolor}{rgb}{0,0.5,0.2}
\definecolor{redcolor}{rgb}{.7,0.,0.}
\definecolor{bluecolor}{rgb}{0,0.,1.}
\definecolor{greycolor}{rgb}{.5,.5,.5}

\usepackage{hyperref}

\usepackage{amsfonts}
\usepackage{url}
\usepackage{units}
\usepackage{array}
\usepackage{eqparbox}
\usepackage[nearskip,caption=false,font=footnotesize]{subfig}
\usepackage{bm}
\usepackage{ulem}
\usepackage{dsfont}

\title{Word Sense Disambiguation Via High Order of Learning in Complex Networks}
\shorttitle{Word Sense Disambiguation Via High Order of Learning in Complex Networks} 

\author{Thiago C. Silva\inst{1} \and Diego R. Amancio\inst{2,*} }
\shortauthor{T. C. Silva and D. R. Amancio}

\institute{

  \inst{1} Institute of Mathematics and Computer Science \\
	University of S\~ao Paulo, P. O. Box 369, Postal Code 13560-970 \\
	S\~ao Carlos, S\~ao Paulo, Brazil \\
    e-mail: thiagoch@icmc.usp.br, thiagochris@gmail.com \\ \ \\

  \inst{2} Institute of Physics of S\~ao Carlos \\
	University of S\~ao Paulo, P. O. Box 369, Postal Code 13560-970 \\
	S\~ao Carlos, S\~ao Paulo, Brazil \\
    e-mail: diego.amancio@usp.br, diegoraphael@gmail.com\\
    $\ast$ Author to whom any correspondence should be addressed. \ \\ \\
}

\pacs{89.75.Hc}{Networks and genealogical trees}
\pacs{89.20.Ff}{Computer science and technology}
\pacs{02.50.Sk}{Multivariate analysis}

\abstract
{
Complex networks have been employed to model many real systems and as a modeling tool in a myriad of applications. In this paper, we use the framework of complex networks to the problem of supervised classification in the word disambiguation task, which consists in deriving a function from the supervised (or labeled) training data of ambiguous words. Traditional supervised data classification takes into account only topological or physical features of the input data. On the other hand, the human (animal) brain performs both low and high level orders of learning and it has facility to identify patterns according to the semantic meaning of the input data. In this paper, we apply a hybrid technique which encompasses both types of learning in the field of word sense disambiguation and show that the high level order of learning can really improve the accuracy rate of the model. This evidence serves to demonstrate that the internal structures formed by the words do present patterns that, generally, cannot be correctly unveiled by only traditional techniques. Finally, we exhibit the behavior of the model for different weights of the low and high level classifiers by plotting decision boundaries. This study helps one to better understand the effectiveness of the model.
}

\begin{document}

\maketitle

\section{Introduction}

Language is present everywhere and pervades all aspects of our daily life since the dawn of humanity. Although it has been largely studied, several issues remain open, such as the explanation of the emergence of fundamental laws such as the Zipf's Law~\cite{manning}. Currently, language has not been exclusively studied by linguists or psychologists. Physicists have borrowed some of their tools to study emergent linguistic patterns. For example, complex systems~\cite{cs}, which are characterized by agents interacting in a non-trivial way, have been used to model interactions between words or segments of a text
~\cite{cancho,eplnovo,eplnovo2}. In the last few years, complex networks (CN) have been used to study both theoretical and practical aspects of language. Examples of recent theoretical findings using such a robust model include the verification of universal properties~\cite{cancho} and the modeling of adjacency
networks. From the practical perspective, complex networks have been used to summarize texts~\cite{summ}, to assess the quality of machine translators~\cite{weaver}, to group and classify data~\cite{thiagoUP,thiagoSSL}, and others.

In the current paper, we assess the ability of complex networks for the Word Sense Disambiguation (WSD) task (i.e., the discrimination of which of the meanings is used in a given context for a word that has multiple meanings). The importance of the WSD task stems from its essential role played for the development of the so called Semantic Web. 
Also, the WSD task is essential for machine translation research~\cite{weaver}. Although a myriad of strategies have been developed so far, none of them evaluated the usefulness of complex networks both to \emph{model texts} and to \emph{recognize patterns} arising from the topological and semantical relationship among senses.
For this reason, we apply a novel generalized methodology based on the concept of complex networks~\cite{thiagoHL} in the field of WSD. First, networks were employed to model the relationship between words in written texts from which it was possible to characterize both the semantical and topological properties of words inserted in a given semantic context (see Section $1.2$ of the Supplementary Information\footnote{The Supplementary Information (SI) is hosted at \url{http://dl.dropbox.com/u/2740286/epl_SI_9apr.pdf}.} (SI)). Then, the similarity relationship given by such a characterization was modeled in the form of networks in order to extract and exploit patterns among the data in the networked representation. Interestingly, assuming that the description of senses in the resulting space is not made up of isolated points, but instead tend to form certain patterns, we found that it is possible to improve the discrimination when we compare the performance achieved with traditional classifiers.

\section{Overview of the Technique}

In this section, we review the hybrid high level technique~\cite{thiagoHL}. Consider a training $X_{training} = \{(x_1,y_1),\ldots,(x_l,y_l)\}$, where the first component of the $i$th tuple $x_i = (f_{1},\ldots,f_{d})$ denotes the attributes of the $d$-dimensional $i$th training instance. The second component $y_{i} \in \mathcal{L} = \{L_1,\ldots,L_n\}$ characterizes the class label or target associated to that training instance. The goal here is to learn a mapping from $x \mapsto y$. Usually, the constructed classifier is checked by using a test set ${X}_{{test}} = \{x_{l+1},\ldots,x_{l+u}\}$, in which labels are not provided. In this case, each data item is called test instance.

In the supervised learning scheme, there are two phases of learning: the \emph{training phase} and the \emph{classification phase}. In the training phase, the classifier is induced or trained by using the training instances (labeled data) in ${X}_{training}$. In the classification phase, the labels of the test instances in $X_{test}$ are predicted using the induced classifier. Below, these two phases are presented in detail.

In the training phase, the data in the training set are mapped into a graph ${G}$ using a network formation technique $g: X_{training} \mapsto G = \langle {V}, {E} \rangle$, where ${V} = \{1,\ldots,V\}$ is the set of vertices and ${E}$ is the set of edges. Each vertex in ${V}$ represents a training instance in ${X}_{{training}}$. As it will be described later, the pattern formation of the classes will be extracted by using the complex topological features of this networked representation.

The edges in ${E}$ are created using a combination of the $\epsilon_r$ and $k$-nearest neighbors ($\kappa$NN) graph formation techniques. In the original versions, the $\epsilon_r$ technique creates a link between two vertices if they are within a distance $\epsilon$, while the $\kappa$NN sets up a link between vertices $i$ and $j$ if $i$ is one of the $k$ nearest neighbors of $j$ or vice versa. Both approaches have their limitations when sparsity or density is a concern. For sparse regions, the $\kappa$NN forces a vertex to connect to its $k$ nearest vertices, even if they are far apart. In this scenario, one can say that the neighborhood of this vertex would contain dissimilar points. Equivalently, improper $\epsilon$ values could result in disconnected components, sub-graphs, or isolated singleton vertices.

The network is constructed using these two traditional graph formation techniques in a combined form. The neighborhood of a vertex $x_{i}$ is given by $N(x_i) = \epsilon_r(x_i, y_{x_i})$, if $|\epsilon_r$($x_{i}$, $y_{x_{i}}| > k$. Otherwise, $N(x_i) = \kappa(x_i, y_{x_i})$, where $y_{x_{i}}$ denotes the class label of the training instance $x_{i}$, \mbox{$\epsilon_r(x_{i}, y_{x_{i}})$} returns the set $\{x_{j}, j \in \mathcal{V} : d(x_{i},x_{j}) < \epsilon \wedge y_{x_{i}} = y_{x_{j}}\}$, and $\kappa(x_{i}, y_{x_{i}})$ returns the set containing the $k$ nearest vertices of the same class as $x_{i}$. Note that the $\epsilon_r$ technique is used for dense regions ($|\epsilon_r(x_{i})| > k$), while the $\kappa$NN is employed for sparse regions. With this mechanism, it is expected that each class will have a unique and single graph component.

\begin{figure} [!Htb]
    \centering
    \subfloat[]
    {\includegraphics[scale=0.16]{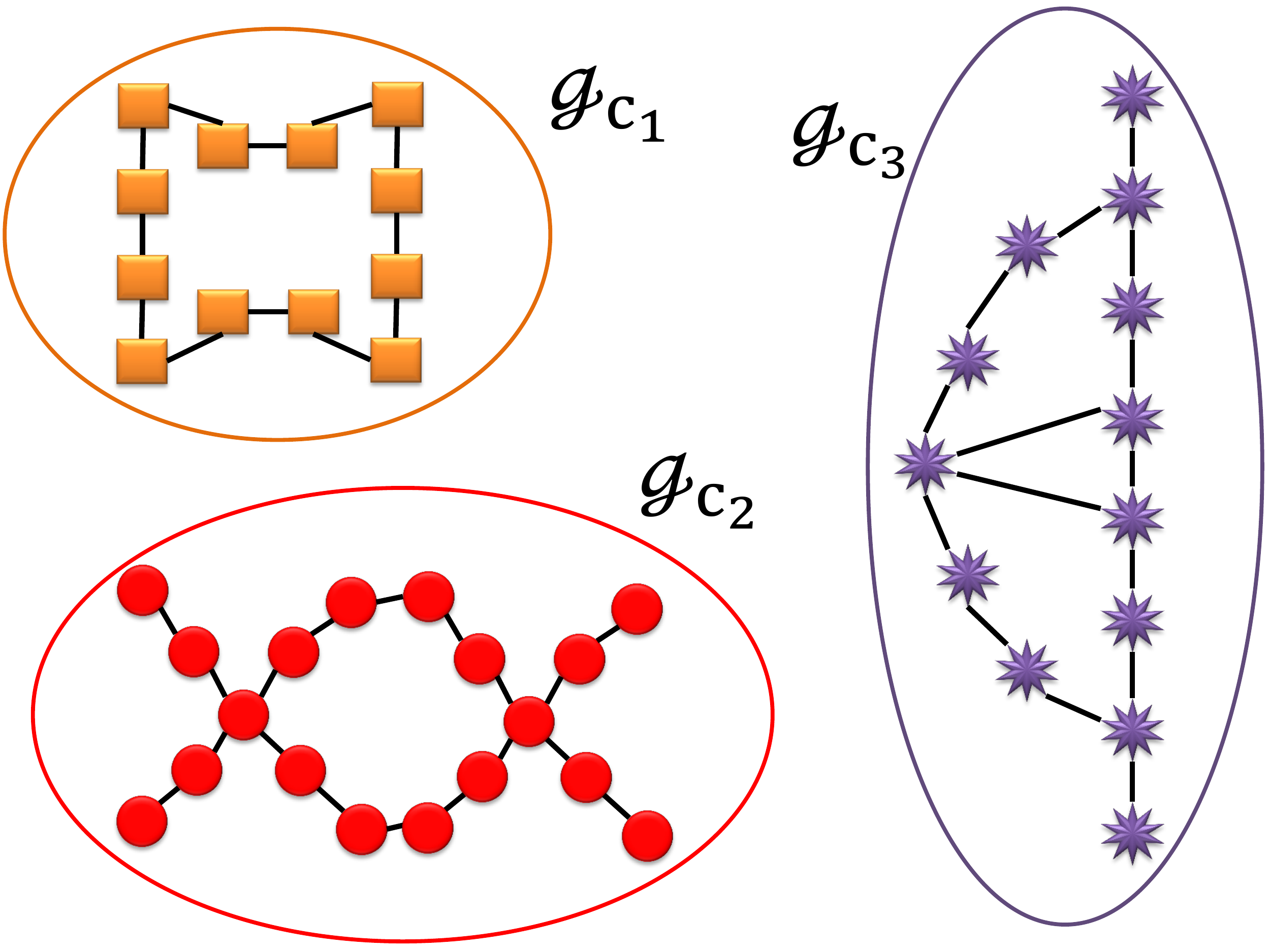}\label{fig:schematic-training}}%
    \subfloat[]
    {\includegraphics[scale=0.16]{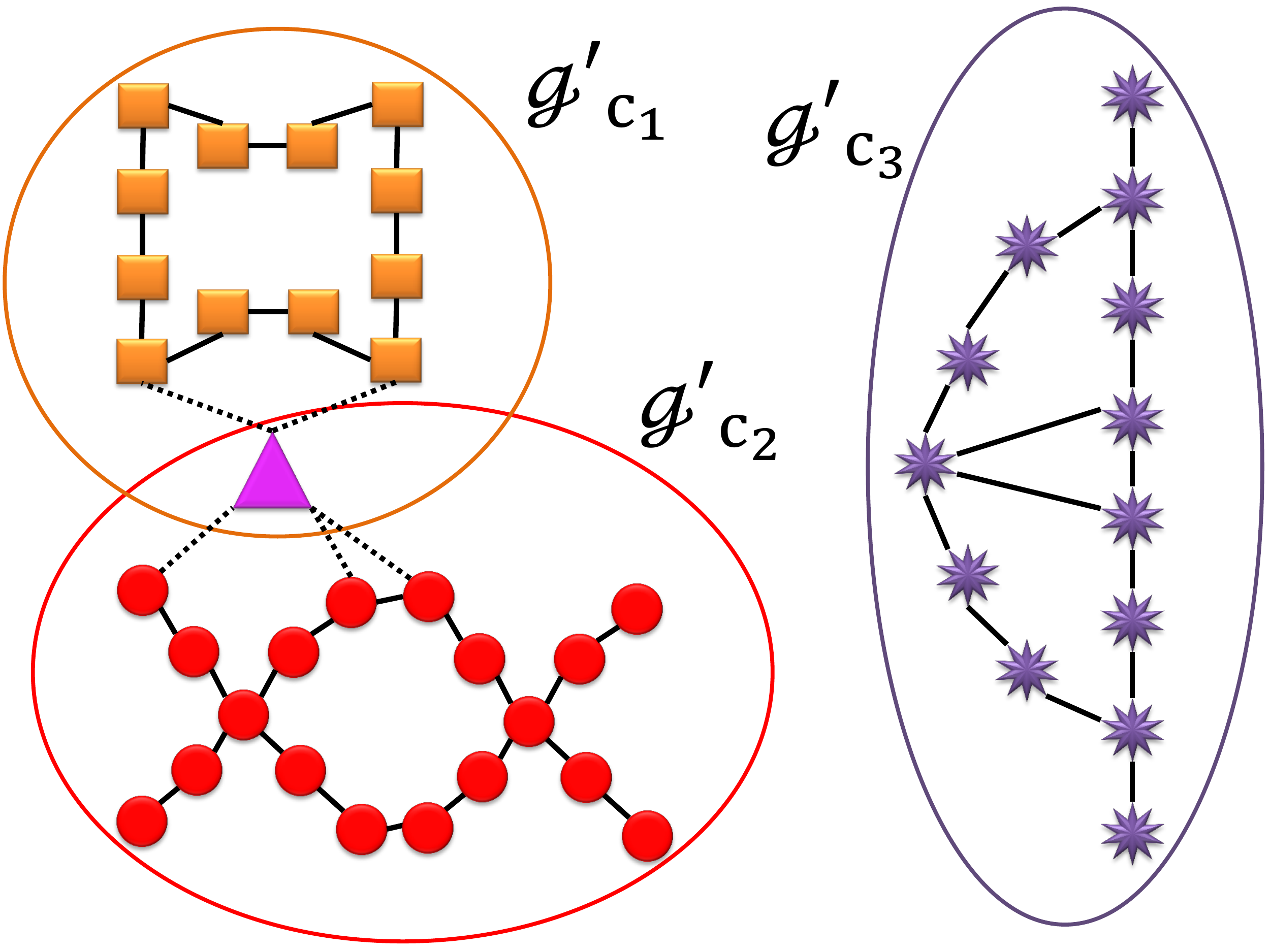}\label{schematic-test}}%
    \caption{(a) Schematic of the network in the training phase.  (b) Schematic of how the classification inference is done.
    }
\end{figure}

For the sake of clarity, Fig. \ref{fig:schematic-training} shows a schematic of how the network looks like for a three-class problem when the training phase has been completed. In this case, each class holds a representative component. In the figure, the surrounding circles denote these components: $\mathcal{G}_{C_{1}}$, $\mathcal{G}_{C_{2}}$, and $\mathcal{G}_{C_{3}}$.

In the classification phase, the unlabeled data items in the $\mathbf{X}_{\mathrm{test}}$ are presented to the classifier one by one. In contrast to the training phase, the class labels of the test instances are unknown. In this way, each test instance is inserted into the network only using the traditional $\epsilon_r$ technique, meaning it is connected to every vertex within this radius, no matter to which class each vertex in this region belongs. Once the data item is inserted, each class analyzes, in isolation, its impact on the respective class component using the complex topological features of it. In the high level model, each class retains an isolated graph component. Each of these components calculate the changes that occur in its pattern formation with the insertion of this test instance. If slight or no changes occur, then it is said that the test instance is in compliance with that class pattern. As a result, the high level classifier yields a great membership value for that test instance on that class. Conversely, if these changes dramatically modify the class pattern, then the high level classifier produces a small membership value on that class. These changes are quantified via network measures, each of which numerically translating the organization of the component from a local to global fashion. As we will see, the average degree, clustering coefficient, and the assortativity measures are employed for the high level order of learning.

For the sake of clarity, Fig. \ref{schematic-test} exhibits a schematic of how the classification process is performed. The test instance (triangle-shaped) is inserted using the traditional $\epsilon_r$ technique. Due to its insertion, the class components become altered: $\mathcal{G}_{C_{1}}^{'}$, $\mathcal{G}_{C_{2}}^{'}$, and $\mathcal{G}_{C_{3}}^{'}$, where each of them is a component surrounded by a circle in Fig. \ref{schematic-test}. It may occur that some class components do not share any links with this test instance. In the figure, this happens with $\mathcal{G}_{C_{3}}^{'}$. In this case, we say that test instance do not comply to the pattern formation of the class component. For the components that share at least a link ($\mathcal{G}_{C_{1}}^{'}$ and $\mathcal{G}_{C_{2}}^{'}$), each of it calculates, in isolation, the impact on its pattern formation by virtue of the insertion of the test instance. For example, when we check the compliance of the test instance to the component $\mathcal{G}_{C_{1}}^{'}$, the connections from the test instance to the component $\mathcal{G}_{C_{2}}^{'}$ are ignored, and vice versa.

Concurrently to the prediction made by the high level classifier, a low level classifier also predicts the membership of the test instance for every class in the problem. The way it predicts depends on the choice of the low level classifier. In the end, the predictions produced by both classifiers are combined via a linear combination to derive the prediction of the high level framework (meta-learning). Once the test instance gets classified, it is either discarded or incorporated to the training set with the corresponding predicted label. In the second case, the classifier must be retrained. Note that, in any of the two situations, each class is still represented by a single graph component.

\section{The High Level Classification}
\label{Classification-Technique}

The hybrid classifier $M$ consists of a convex combination of two terms: (i) a low level classifier (C4.5~\cite{bishop}, kNN~\cite{bishop} or Naive Bayes~\cite{bishop})\footnote{A brief description of the low level classifiers is given in the SI.}; and (ii) a high level classifier, which is responsible for classifying a test instance according to its pattern formation with the data. Mathematically, the membership of the test instance $x_{i} \in \mathbf{X}_{\mathrm{test}}$ with respect to the class $j \in \mathcal{L}$, here written as $M^{(j)}_{i}$, is given by:

\begin{equation}
    M^{(j)}_{i} = (1 - \lambda)T^{(j)}_{i} + \lambda C^{(j)}_{i},
    \label{eq:def-classification}
\end{equation}

\noindent where $T^{(j)}_{i} \in [0,1]$ denotes the membership of the test instance $x_{i}$ on class $j$ produced by an arbitrary traditional (low level) classifier; $C^{(j)}_{i} \in [0,1]$ indicates the same membership information yielded by a high level classifier; and $\lambda \in [0,1]$ is the \emph{compliance term}, which plays the role of counterbalancing the classification decision supplied by both low and high level classifiers. Whenever $T^{(j)}_{i} = 1$ and $C^{(j)}_{i} = 1$, we may deduce that the $i$th data item carries all the characteristics of class $j$. On the other hand, whenever $T^{(j)}_{i} = 0$ and $C^{(j)}_{i} = 0$, we may infer that the $i$th data item does not present any similarities nor complies to the pattern formation of class $j$. Values in-between these two extremes lead to natural uncertainness in the classification process and are found in the majority of times during a classification task. Note that Eq. (\ref{eq:def-classification}) generates fuzzy outputs. Moreover, it is valuable to indicate that, when $\lambda = 0$, Equation (\ref{eq:def-classification}) reduces to a common low level classifier. A test instance receives the label from the class $j$ that maximizes (\ref{eq:def-classification}).

The inference of pattern formation, which is used by the classifier $C$, within the data is processed using the generated network. The motivation behind using networks is that it can describe topological structures among the data items. These networks are constructed such that: (i) each class is an isolated subgraph (component) and (ii) after the insertion of a new test instance, each class must still retain a representative and unique component. With that in mind, the pattern formation of the data is quantified through a combination of network measures developed in the complex network literature. These measures are chosen in a way to cover relevant high level aspects of the class component.
Suppose that $K$ measures are selected to comprise the high level classifier $C$. Mathematically, the membership of the test instance $x_{i} \in \mathbf{X}_{\mathrm{test}}$ with respect to the class $j \in \mathcal{L}$ yielded by the high level classifier, here written as $C^{(j)}_{i}$, is given by:

\begin{equation}
    C^{(j)}_{i} = \frac{\sum_{u = 1}^{K}{\alpha(u)\left[1 - f^{(j)}_{i}(u) \right]}}{\sum_{g \in L}{\sum_{u = 1}^{K}{\alpha(u)\left[1 - f^{(g)}_{i}(u) \right]}}},
    \label{eq:def-C-term}
\end{equation}

\noindent where $\alpha(u) \in [0,1], \forall u \in \{1,\ldots,K\}$, $\sum_{u = 1}^{K}{\alpha(u)} = 1$, are user-controllable coefficients that indicate the influence of each network measure in the classification process and $f^{(j)}_{i}(u)$ is a function that depends on the $u$th network measure applied to the $i$th data item with regard to the class $j$. This function is responsible for providing an answer whether the test instance $x_{i}$ presents the same patterns of the class $j$ or not. The denominator in (\ref{eq:def-C-term}) has been introduced solely for normalization matters.

With respect to $f^{(j)}_{i}(u)$, it possesses a general closed form given by:

\begin{equation}
    f^{(j)}_{i}(u) = \Delta G^{(j)}_{i}(u) p^{(j)},
    \label{eq:f-function-def}
\end{equation}

\noindent where $\Delta G^{(j)}_{i}(u) \in [0,1]$ is the variation of the $u$th network measure that occurs on the component representing class $j$ if $x_{i}$ joins it and $p^{(j)} \in [0,1]$ is the proportion of data items pertaining to the class $j$. Remembering that each class has a component representing itself, the strategy to check the pattern compliance of a test instance is to examine whether its insertion causes a great variation of the network measures representing the class component. In other words, if there is a small change in the network measures, the test instance is in compliance with all the other data items that comprise that class component, i.e., it follows the same pattern as the original members of that class. On the other hand, if its insertion is responsible for a significant variation of the component's network measures, then probably the test instance may not belong to that class.

We proceed to explain the role of the $p^{(j)} \in [0,1]$ in (\ref{eq:f-function-def}). In real-world databases, unbalanced classes are usually encountered. In general, a database frequently encompasses several classes of different sizes. A great portion of the network measures are very sensitive to the size of the components. In an attempt to soften this problem and cancel out the effects of distinct components' sizes, (\ref{eq:f-function-def}) introduces the term $p^{(j)}$, which is the proportion of vertices that class $j$ has.

\subsection{Composition of the High Level Classifier}

The network measurements that compose the high level classifier are the assortativity~\cite{paternos}, the clustering coefficient, and the average degree. The reason why these three measures have been chosen is as follows: the average degree measure figures out strict local scalar information of each vertex in the network; the clustering coefficient of each vertex captures local structures by means of counting triangles formed by the current vertex and any of its two neighbors; the assortativity coefficient considers not only the current vertex and its neighbors, but also the second level of neighbors (neighbor of neighbor), the third level of neighbors, and so on. We can perceive that the three measures characterize the network's topological properties in a local to global fashion. In this way, the combination of these measures is expected to capture the pattern formation of the underlying network in a systematic manner. Details regarding these three measurements are given in the SI.

\section{Results and Discussion}

First, the methodology is applied to an artificial database in order to better understand its functionality. Afterwards, the WSD problem is analyzed. The discussion of the observed results is given below.

\subsection{High Level Applied to a Toy Database}

\begin{figure*} [!Htb]
    \centering
    \subfloat[]
    {\includegraphics[width=0.33\textwidth]{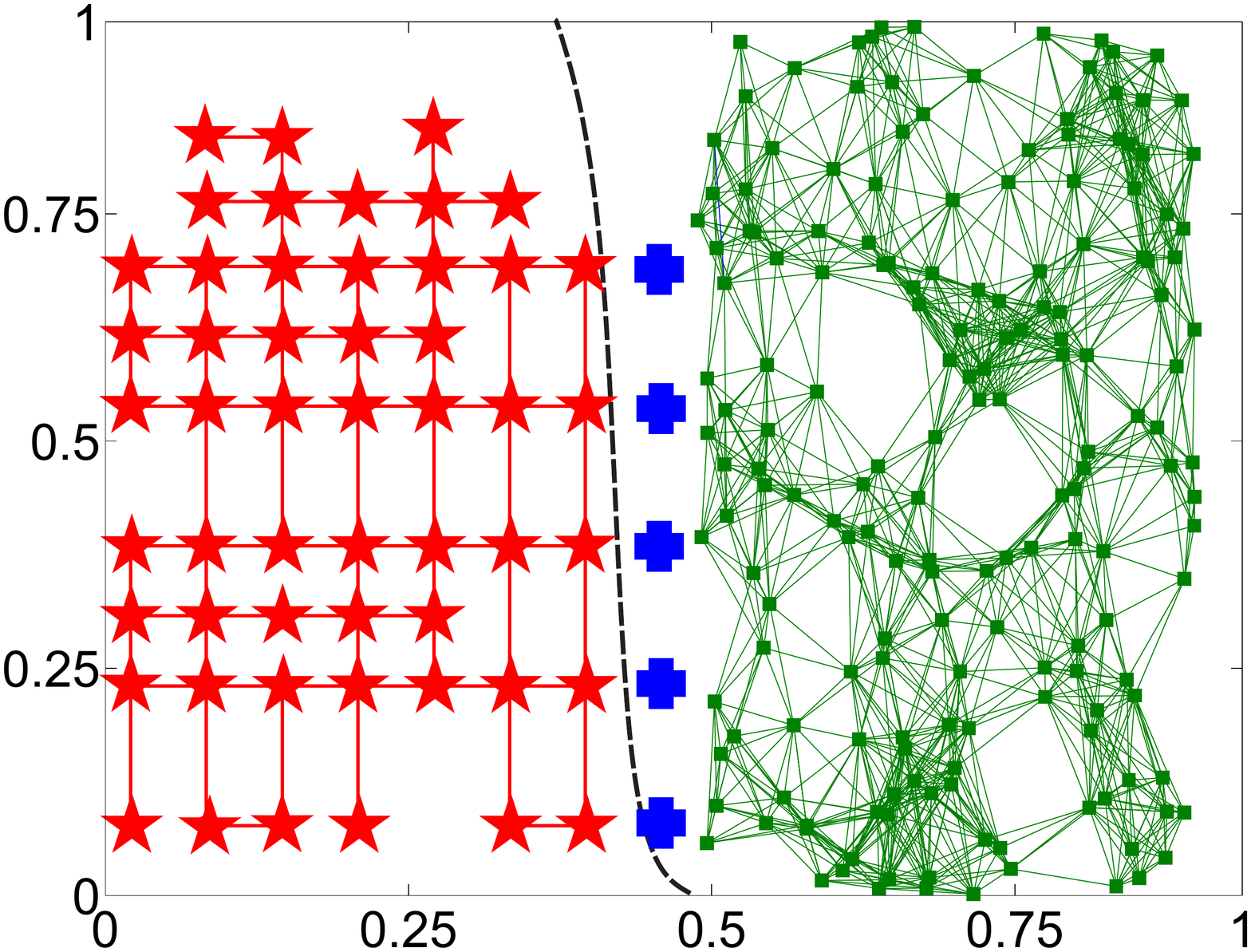}\label{fig:Regular-Lattice-lambda0}}%
    \subfloat[]
    {\includegraphics[width=0.33\textwidth]{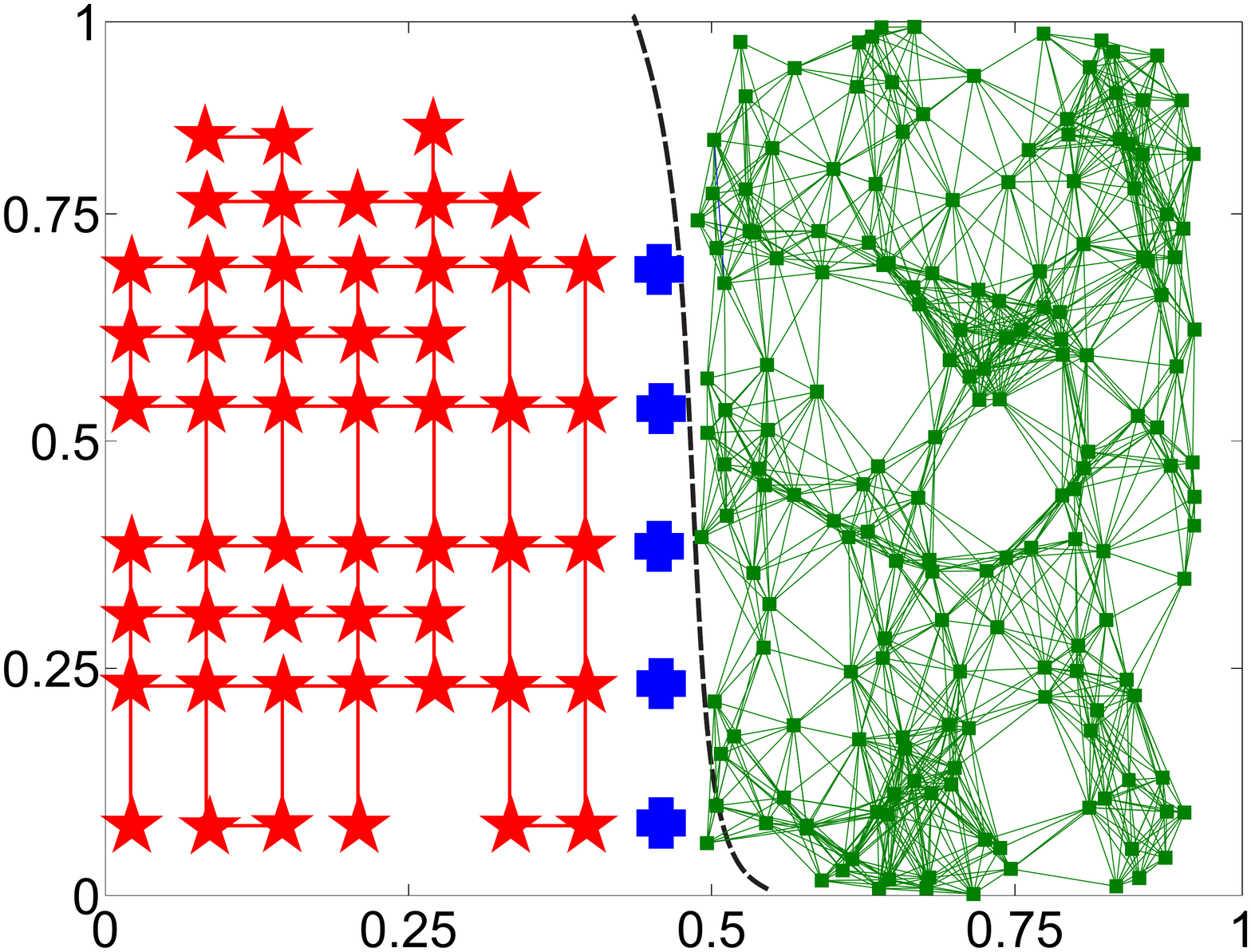}\label{fig:Regular-Lattice-lambda05}}%
    \subfloat[]
    {\includegraphics[width=0.33\textwidth]{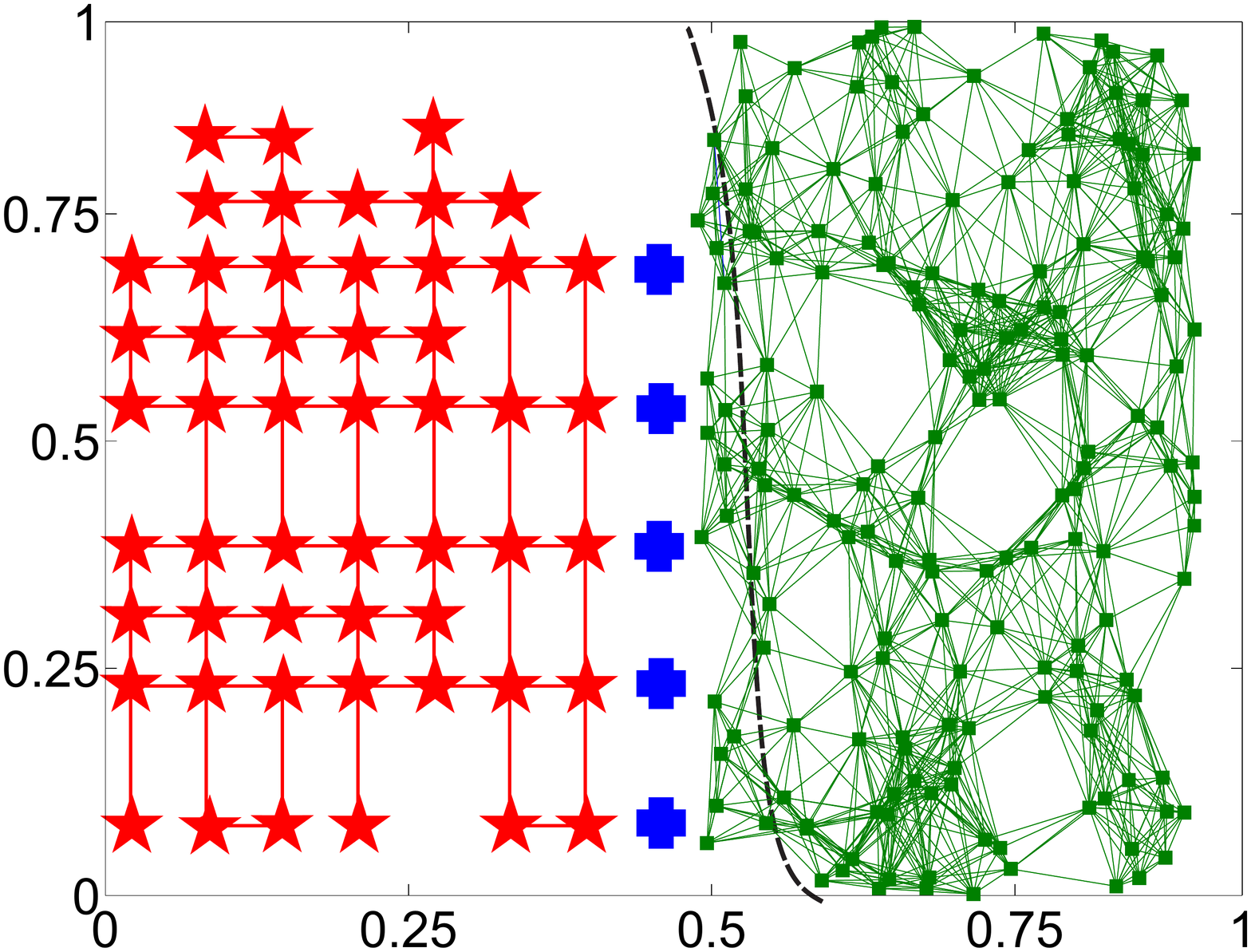}\label{fig:Regular-Lattice-lambda08}}%
    \caption{\label{full}Behavior of the decision boundaries as $\lambda$ varies in the toy data. Decision boundaries when (a) $\lambda = 0$; (b) $\lambda = 0.5$; and (c) $\lambda = 0.8$.
    }
\end{figure*}

As an introductory example, consider the toy data set depicted in Fig. 
\ref{full},
where there are two classes: the red or ``star" ($52$ vertices) and the green or ``square" ($276$ vertices) classes. This example serves as a gist of how the hybrid classifier draws its decisions. In the training and classification phases, we employ $\kappa = 3$ and $\epsilon = 0.04$ for the network construction. The fuzzy SVM~\cite{svmref} with RBF kernel ($C=70$ and $\gamma=2^{-1}$) is adopted for the low level classifier.  By inspection of the figure, the red or ``star" class displays a well-defined pattern: a grid or lattice, whereas the green or ``square" class does not indicate any well-established patterns. Here, the goal is to classify the cross-shaped data items (test set) one by one using only the information of the training set. Figures \ref{fig:Regular-Lattice-lambda0}, \ref{fig:Regular-Lattice-lambda05}, and \ref{fig:Regular-Lattice-lambda08} exhibit the decision boundaries of the two classes when $\lambda = 0$, $\lambda = 0.5$, and $\lambda = 0.8$, respectively. When $\lambda = 0$, only the SVM prediction is used by the hybrid technique. In this case, one can see that the five data items are not correctly classified. Notice that the decision boundaries are pushed near the red or ``star'' class by virtue of the large amount of green or ``square'' items in the vicinity. Now, when $\lambda = 0.5$, the SVM and the high level classifier predictions are utilized in the same intensity. In this situation, the decision boundaries are dragged toward the green or ``square'' class, because of the strong pattern that the red or ``star'' class exhibits. We can think this phenomenon as being a clash between the two decision boundaries: as $\lambda$ increases, the more structured class tends to possess more decision power, and, consequently, is able to reduce the effective area of the competing class. For example, when $\lambda = 0.8$, the organizational features of the red or ``star'' class are so salient that its effective area invades the high density region of the green or ``square'' class. In the two former cases, the hybrid high level technique can successfully classify the cross-shaped data items. In summary, the concept of classification is altered depending on the value of the compliance term. A small compliance term causes the final decision of the hybrid classifier to be rooted in traditional assumptions of low level classifiers. Now, when a large compliance term is used, the salient characteristic that the hybrid classifier attempts to emphasize is the patterns that the classes display. As the structural pattern of a class becomes stronger, wider will be the delineated decision boundary for that class.


\subsection{High Level Applied to Word Sense Disambiguation}

    The efficiency of the high level classifier is also verified in a real-world application. In this case, we aim at discriminating senses of ambiguous words (i.e., words with the same lexical form but with different senses)\footnote{For example, the word ``bear'' might be either related to a large mammal of the family {\ it Ursidae} or to the verb ``carry''.}. Using the database presented in Ref~\cite{amancioEPL}, two approaches for characterizing senses are employed: the topological and the semantical approach. In the former, each occurrence of a word is characterized by its local structure in the word adjacency network~\cite{amancio2}. In the latter, each word sense is represented by the frequency of the $w$ nearby words. Details of these two methodologies are given in the SI.

    Table \ref{tab.1} shows the results obtained for the five ambiguous words in the topological approach. Similarly, Table \ref{tab.2} depicts the results obtained for the semantic approach. In both cases, when selecting the suitable value of the parameter $\lambda$, it is possible to improve the efficiency of the classification achieved by the low level classifiers (C4.5, kNN and Naive Bayes). Moreover, because $\lambda$ is different from zero in most cases, one can infer that there is a pattern in the data organized in the attribute space. Interestingly, one can conclude that the structural organization of the words in complex networks is not only useful for discriminating senses when modeling the relationship of words in a text, but also when modeling the relationship between words in the attribute space. In other words, when word senses are analyzed with the complex network framework, patterns emerge both in the organization of words in the adjacency network adjacency (before characterization) and in the network built in the attribute space. These unveiled patterns, in turn, cannot be properly discovered by traditional techniques. This reasoning explains the performance boost that occurred when a $\lambda \neq 0$ was employed in the experiments.

\begin{table*}
        \centering
        \caption{\label{tab.1} Structural approach for discriminating senses of ambiguous words. Senses were characterized according to topological CN measurements~\cite{amancioEPL} and the discrimination of senses was performed with low (kNN, C4.5 and Bayes) and high level classifiers.  Note that the high level technique always outperforms the traditional low level classification.}
            \begin{tabular}{|c|c|c|c|c|c|c|}
            \hline
            Approach      & Low Level  & \multicolumn{2}{|c|}{Low Level Classification} & \multicolumn{3}{|c|}{High Level Classification}                \\
            \hline
            \textbf{Word} & \textbf{Algorithm} & \textbf{Acc. Rate} & \textbf{p-value} & \textbf{Acc. Rate} & \textbf{p-value} & \textbf{Best $\lambda$}  \\
            \hline
                            & kNN  & 87.6 \%  &  $1.9 \times 10^{-4}$ & 90.2 \% & $4.9 \times 10^{-6}$ & 0.10 \\
            save            & C4.5 & 79.8 \%  &  $2.1 \times 10^{-1}$ & 84.2 \% & $9.7 \times 10^{-3}$ & 0.25 \\
                            & Bayes& 83.1 \%  &  $2.5 \times 10^{-2}$ & 86.5 \% & $8.4 \times 10^{-4}$ & 0.15 \\
            \hline
                            & kNN  & 84.5 \%  &  $1.8 \times 10^{-1}$ & 85.3 \% & $1.3 \times 10^{-1}$ & 0.10 \\
            note            & C4.5 & 78.4 \%  &  $8.7 \times 10^{-1}$ & 80.3 \% & $7.0 \times 10^{-1}$ & 0.15 \\
                            & Bayes& 78.4 \%  &  $8.7 \times 10^{-1}$ & 80.3 \% & $7.0 \times 10^{-6}$ & 0.15 \\
            \hline
                            & kNN  & 87.0 \%  &  $4.3 \times 10^{-3}$ & 88.4 \% & $9.9 \times 10^{-4}$ & 0.10 \\
            march           & C4.5 & 60.9 \%  &  $5.8 \times 10^{-1}$ & 71.9 \% & $1.7 \times 10^{-1}$ & 0.35 \\
                            & Bayes& 73.9 \%  &  $1.7 \times 10^{-1}$ & 76.2 \% & $8.7 \times 10^{-2}$ & 0.20 \\
            \hline
                            & kNN  & 71.1 \%  &  $1.6 \times 10^{-3}$ & 71.1 \% & $1.6 \times 10^{-3}$ & 0.00  \\
            present         & C4.5 & 64.7 \%  &  $2.0 \times 10^{-1}$ & 65.5 \% & $1.4 \times 10^{-1}$ & 0.05  \\
                            & Bayes& 73.9 \%  &  $4.9 \times 10^{-5}$ & 73.9 \% & $4.9 \times 10^{-2}$ & 0.00  \\
            \hline
                            & kNN & 100.0 \% &  $6.0 \times 10^{-3}$ & 100.0 \% &  $6.0 \times 10^{-3}$ & 0.00  \\
            jam             & C4.5 & 80.0 \% &  $1.7 \times 10^{-1}$ & 84.4 \%  &  $1.6 \times 10^{-1}$ & 0.20  \\
                            & Bayes& 90.0 \% &  $4.6 \times 10^{-2}$ & 91.9 \%  &  $4.6 \times 10^{-2}$ & 0.10  \\
            \hline
            \end{tabular}
        \end{table*}

 \begin{table*}
        \centering
        \caption{\label{tab.2} Semantic approach for discriminating senses of ambiguous words. Senses were characterized according to frequency of the $n=5$ neighbors of the ambiguous word~\cite{amancioEPL} and the discrimination of senses was performed with low (kNN, C4.5 and Bayes) and high level classifiers. Acc. Rate represents the accuracy rate obtained with an evaluation based on the 10-fold cross-validation technique~\cite{validation}. The p-value refers to the likelihood of obtaining the same accuracy rate with an random classifier (see Ref.~\cite{amancioEPL} for details). Note that the high level technique always outperforms the traditional low level classification.}
            \begin{tabular}{|c|c|c|c|c|c|c|}
            \hline
            Approach      & \multicolumn{3}{|c|}{Low Level Classification} & \multicolumn{3}{|c|}{High Level Classification}               \\
            \hline
            \textbf{Word} & \textbf{Algorithm} &\textbf{Acc. Rate} & \textbf{p-value} & \textbf{Acc. Rate} & \textbf{p-value} & \textbf{Best $\lambda$}  \\
            \hline
                           & kNN   & 79.5 \%  & $6.2 \times 10^{-1}$  & 84.7 \% & $7.2 \times 10^{-2}$ & 0.25 \\
            save           & C4.5  & 78.9 \%  & $6.9 \times 10^{-1}$  & 84.1 \% & $1.0 \times 10^{-1}$ & 0.25 \\
                           & Bayes & 76.6 \%  & $8.9 \times 10^{-1}$  & 81.9 \% & $3.2 \times 10^{-1}$ & 0.25 \\
            \hline
                           & kNN   & 82.6 \%  & $3.8 \times 10^{-1}$  & 86.4 \% & $6.1 \times 10^{-1}$ & 0.20\\
            note           & C4.5  & 79.5 \%  & $7.6 \times 10^{-1}$  & 83.1 \% & $3.1 \times 10^{-1}$ & 0.15\\
                           & Bayes & 78.3 \%  & $8.6 \times 10^{-1}$  & 82.1 \% & $4.7 \times 10^{-1}$ & 0.20\\
            \hline
                           & kNN   & 82.8 \%  & $1.4 \times 10^{-2}$  & 89.7 \% & $9.9 \times 10^{-4}$ & 0.30\\
            march          & C4.5  & 82.8 \%  & $1.4 \times 10^{-2}$  & 87.4 \% & $4.3 \times 10^{-3}$ & 0.30\\
                           & Bayes & 62.5 \%  & $5.8 \times 10^{-1}$  & 72.9 \% & $1.7 \times 10^{-1}$ & 0.35\\
            \hline
                           & kNN   & 62.9 \%  & $4.2 \times 10^{-1}$ & 69.4 \% & $7.9 \times 10^{-3}$  & 0.20\\
            present        & C4.5  & 57.0 \%  & $9.6 \times 10^{-1}$ & 61.4 \% & $6.3 \times 10^{-1}$  & 0.15\\
                           & Bayes & 60.2 \%  & $7.6 \times 10^{-1}$ & 65.3 \% & $1.7 \times 10^{-1}$  & 0.20\\
            \hline
                           & kNN   & 76.5 \%  & $6.3 \times 10^{-1}$ & 87.3 \% & $4.6 \times 10^{-2}$ & 0.35\\
            jam            & C4.5  & 76.5 \%  & $6.3 \times 10^{-1}$ & 89.5 \% & $4.4 \times 10^{-2}$ & 0.40\\
                           & Bayes & 82.4 \%  & $4.1 \times 10^{-1}$ & 90.1 \% & $4.1 \times 10^{-2}$ & 0.30\\
            \hline
            \end{tabular}
        \end{table*}

    \section{Conclusion}

    In the current paper, we have applied a novel methodology of supervised data classification in the field of word sense disambiguation. The hybrid classifier is comprised of a combination of traditional (low level) and pattern-based classifiers. The latter uses a network to exploit the topological patterns in search of patterns. From the analysis of the experiments, we have found that the inclusion of the high level term was responsible for improving the ability of classification both in artificial and real-world networks. Specifically, in the latter, the methodology devised in Ref.~\cite{amancioEPL} was improved as a consequence that words conveying the same meaning display organizational patterns not only in textual level but also in the attribute space. This argument serves to strengthen the fact that networks constructed using words are not totally disorganized. Instead, each set of words tend to form patterns that uniquely describe it. The hybrid framework exactly attempts to extract these hidden patterns that are cloaked within the word relationships (edges) in the network.

    Because the hybrid high level technique is totally generic, we intend to use it in other real-world applications, other than word disambiguation. In addition, a methodology for automatically finding the best value of the compliance term will also be the subject of our future studies.

    \acknowledgments
    TCS (2009/12329-1) and DRA (2010/00927-9) acknowledge the financial support from FAPESP.

\end{document}